\documentclass[11pt,a4paper]{article}
\pdfoutput=1

\usepackage{amsmath}
\usepackage{amsfonts}
\usepackage{amsthm}
\usepackage{amssymb}
\usepackage{amscd}
\usepackage[british]{babel}
\usepackage{graphicx}
\usepackage{psfrag}
\usepackage{epsfig}
\usepackage{rotating}
\usepackage{times}

\theoremstyle{plain}

\newtheorem{remark}{Remark}[section]

\newcommand{\boxend}{\flushright{$\Box$}}

\begin{document}

\title{Future singularity avoidance in phantom dark energy models}

\author{Jaume de Haro$^{a,}$\footnote{E-mail: jaime.haro@upc.edu}
 %and Emilio Elizalde$^{b,}$\footnote{E-mail: elizalde@ieec.uab.es, elizalde@math.mit.edu}
 }

\maketitle

{$^a$Departament de Matem\`atica Aplicada I, Universitat
Polit\`ecnica de Catalunya, Diagonal 647, 08028 Barcelona, Spain
%\\
%$^b$Instituto de Ciencias del Espacio (CSIC) \& Institut
%d'Estudis Espacials de Catalunya (IEEC/CSIC)\\ Campus UAB, Facultat
%de Ci\`encies, Torre C5-Parell-2a planta, 08193 Bellaterra
%(Barcelona) Spain
}

\thispagestyle{empty}

\begin{abstract}
Different approaches to quantum cosmology are studied in order to deal with the future singularity avoidance problem. Our results show that
these future singularities will persist but  could take different forms. As an example we have studied the big rip  which appear when one 
considers the state equation $P=\omega\rho$
with $\omega<-1$, showing that it does not disappear in
modified gravity. On the other hand, it is well-known that quantum geometric effects (holonomy corrections) in loop quantum 
cosmology introduce a quadratic modification, namely proportional to $\rho^2$, in Friedmann's equation
that replace the big rip by a non-singular bounce. However this modified Friedmann equation could have been obtained in an inconsistent way, what
means that the obtained results from this equation, in particular singularity avoidance, would be incorrect. In fact, we will show that instead of 
a non-singular bounce, the big rip
singularity would be replaced, in loop quantum cosmology, by other kind of
 singularity.
\end{abstract}

{\bf Pacs numbers:} 98.80.Qc, 04.20.Dw, 04.62.+v

%\maketitle

\vspace{0.5cm}

\noindent{\it 1.- Introduction}---
Studies of distant type Ia
supernovae \cite{p99,r99}
indicates that  the dominant part of the energy of the universe must be gravitationally repulsive driving  our
universe  expanding in an accelerating way. To explain this
acceleration one usually assumes the existence of dark energy with a
negative pressure, in general one can assume a perfect fluid with state equation $P=\omega \rho$, with $\omega<-1/3$ in order 
to have cosmic acceleration.
Moreover, observations from WMAP indicates the value $\omega\cong-1.10$ \cite{k11}, what means that our universe
would be dominated by ``phantom energy`` ($\omega<-1$). However, the classical solutions of general
relativity for a Friedmann-Robertson-Walker (FRW) model containing
dark energy lead, in general, to future singularities
\cite{ckw03,b04,not05} (big rip, future sudden singularities,
etc.). Lately, a good number of papers have been dealing with the
possibility of avoiding these future
singularities, using  different approaches to quantum cosmology like loop quantum cosmology, semiclassical gravity, modified gravity, 
brane cosmology, etc.
This paper has two main objectives: The first one is to discuss this different approaches, to show  in which way they modify the dynamics 
of our universe
and to check if, effectively, they could avoid the future singularities that appear in classical cosmology.
And the second one is to show that  the modified Friedmann equation in loop quantum cosmology could have been obtained in an inconsistent way. 
And thus, the current statement that, in loop quantum cosmology,
 the big rip singularity is
replaced by a non-singular bounce would be incorrect.

 The units used in this paper are
$\hbar=c=M_p=1$ being $M_p$ the reduced Planck mass.

\vspace{0.5cm}

\noindent{\it 2.- Einstein cosmology}---
For the flat FRW spacetime  filled by a perfect fluid with state equation $P=f(\rho)$, Einstein theory, is obtained from the Lagrangian
${\mathcal L}=\frac{1}{2}Ra^3-\rho a^3$  where $R=6(\dot{H}+2H^2)$ is the scalar curvature, $a$ is the
scale factor and $H=\frac{\dot{a}}{a}$ is the Hubble parameter.

This Lagrangian has been constructed in co-moving fluid coordinates (see \cite{s03} and Section III C of  \cite{r72}), and
the energy density $\rho$ has to be understood as a function of the scalar factor $a$. This relation comes from the
consevation equation 
\begin{eqnarray}\label{2}
 d(\rho a^3)=-Pd(a^3)\Longleftrightarrow
\dot{\rho}=-3H(\rho+P)\Longleftrightarrow \frac{d\rho}{da}=-\frac{3}{a}(\rho+P).
\end{eqnarray}

Since the state equation that we are studying is $P=f(\rho)$ one has the differential equation
\begin{eqnarray}
 \frac{d\rho}{\rho+f(\rho)}=-\frac{3}{a}da,
\end{eqnarray}
that after integration gives $\rho$ as a function of $a$. For example, when $P=\omega \rho$ one has 
\begin{eqnarray}
 \rho(a)=\rho_0\left(\frac{a}{a_0}\right)^{-3(1+\omega)},
\end{eqnarray}
where $\rho_0$ is the value of $\rho$ when $a=a_0$.

The Lagrangian can be written as follows ${\mathcal L}=3\left(\frac{d{(\dot{a} a^2)}}{dt}-\dot{a}^2 a\right)-\rho a^3$, this means that the same theory
is obtained avoiding the total derivative, which gives the Lagrangian ${\mathcal L}_{E}=-3H^2 a^3-\rho a^3$. The conjugate momentum is  then given by
$p=\frac{\partial {\mathcal L}_{E}}{\partial\dot{a}}=-6Ha^2$, and thus the Hamiltonian is
\begin{eqnarray}\label{1}
{\mathcal H}_{E}=
\dot{a}p- {\mathcal L}_{E}= -3H^2a^3+\rho a^3.\end{eqnarray}

In general relativity  the Hamiltonian is constrained to be zero,
which gives the
Friedmann equation
\begin{eqnarray}
H^2=\rho/3,
\end{eqnarray}
that together  with the conservation equation
are the dynamical equations that describe the evolution of the universe.

 The Raychaudury equation is obtained from the Hamilton equation
$\dot{p}=-\frac{\partial {\mathcal H}_E}{\partial a}$, which
gives
\begin{eqnarray}\label{3}
\dot{p}= -\frac{p^2}{12 a^2}-\frac{\partial (\rho a^3)}{\partial a}= -\frac{p^2}{12 a^2}+3 P a^2,\end{eqnarray}
where we have used the conservation equation. Then from the Friedmann equation one easily obtains $\dot{H}=-\frac{1}{2}(\rho+P)$.

A solvable example is the case of a barotropic perfect fluid with state equation $P=\omega\rho$ \cite{ckw03}, which gives
\begin{eqnarray}\label{a1}
{H}(t)=\frac{2}{3(1+\omega)}\frac{1}{t-t_s} \quad \rho(t)=\frac{4}{3(1+\omega)^2}\frac{1}{(t-t_s)^2},
\end{eqnarray}
where $t_s\equiv t_0-\frac{2}{3H_0(1+\omega)}$, being $H_0=H(t_0)$
the initial condition.
Then, if one assumes $H_0>0$ and $\omega<-1$
one has $t_s>t_0$, and thus, one has a big rip  singularity.

\vspace{0.5cm}

\noindent{\it 3.- Modified gravity}---
An alternative  to Einstein cosmology is modified gravity, where
higher-curvature terms are taken into account. The theory is
 based in the Lagrangian
(see for instance \cite{apt07,no03,bno08})
${\mathcal L}_{MG}=f(R)a^3-a^3\rho$, and  to find the Hamiltonian formulation, one can use the
Ostrogradskii's construction  \cite{me07}
introducing the variables
$a_1\equiv a$ and $a_2\equiv \dot{a}$, and  the momenta
\begin{eqnarray}\label{15}
p_1\equiv\frac{\partial {\mathcal L}_{MG}}{\partial\dot{a}}-\frac{d}{dt}\frac{\partial {\mathcal L}_{MG}}{\partial\ddot{a}}=-6a^2f''(R)\dot{R},
\quad
p_2\equiv \frac{\partial {\mathcal L}_{MG}}{\partial\ddot{a}}=6a^2f'(R).
\end{eqnarray}

Then,
the Hamiltonian in modified gravity is given by
\begin{eqnarray}\label{17}
{\mathcal H}_{MG}\equiv p_1\dot{a}+p_2\ddot{a}-{\mathcal L}_{MG}=
\nonumber\\ 
\left(-6f''(R)\dot{R}H+f'(R)(R-6H^2)-f(R)+\rho \right)a^3,
\end{eqnarray}
and the Hamiltonian constraint ${\mathcal H}_{MG}=0$ gives the modified Friedmann equation
\begin{eqnarray}\label{18}
-6f''(R)\dot{R}H+f'(R)(R-6H^2)-f(R)+\rho=0,
\end{eqnarray}
that with the conservation equation $\dot{\rho}=-3H(P+\rho)$ gives  the dynamics of the universe in modified gravity.

Note that, the Hamilton equations $\dot{a}_1=\frac{\partial{\mathcal H}_{MG}}{\partial p_1}$, $\dot{a}_2=\frac{\partial{\mathcal H}_{MG}}{\partial p_2}$
and $\dot{p}_2=-\frac{\partial{\mathcal H}_{MG}}{\partial a_2}$ are identities. The dynamical equation, i.e. the Euler-Lagrange equation, is
 $\dot{p}_1=-\frac{\partial{\mathcal H}_{MG}}{\partial a_1}$ which gives the modified Rauchaudury equation.

\begin{remark}
 A more general theory consist in modified Gauss-Bonet gravity which is based on the Lagrangian ${\mathcal L}_{GB}=f(R,G)a^3-a^3\rho$, being
$G=24H^2(\dot{H}+H^2)$ the scalar Gauss-Bonet curvature (see for instance \cite{cenoz06}).
\end{remark}

As special case we can consider $f(R)=\frac{R}{2}-\frac{\alpha}{12}R^2$. Then, equation (\ref{18}) becomes
\begin{eqnarray}\label{19}
H^2=\frac{\rho}{3}+2\alpha(3H^2\dot{H}+H\ddot{H}-\frac{1}{2}\dot{H}^2),
\end{eqnarray}
which coincides with the semiclassical Friedmann equation, obtained taking into account quantum effects due to a massless conformally coupled field
when one chooses the other  parameter, namely $\beta$, equal to zero (see for instance \cite{hae11}).

For the state equation $P=\omega\rho$, the case $\omega<-1, \alpha>0$   was studied in great detail in \cite{hae11}. There, the main  obtained result
is that almost all the solution have future singularities in the contracting phase, having the following behavior:
\begin{eqnarray}\label{20}
 H(t)\sim -\frac{1}{2(t_s-t)},\quad \rho(t)\sim 0, \quad \mbox{for} \quad t<t_s
\end{eqnarray}
 being $t_s$ the time at which the future singularity appears. This behavior means that if the universe is initially in the expanding phase, like
nowadays, it will bounce and  will enter in the contracting phase where it will develop the singularity described by (\ref{20}).

On the other hand,
the case $\omega<-1, \alpha<0$ is completely different, because now the expanding and contracting phase decouple, that is, the bounces are
not allowed. To show this, we consider the new variable $\bar{p}^2=\epsilon H$ where $\epsilon=sign(H)$. Using this variable the 
modified Friedmann equation becomes
 \begin{eqnarray}\label{21}
\frac{d}{d{t}}\left(\dot{\bar p}^2/2+{V}(\bar{p},{\rho})\right)=-3\epsilon
\bar{p}^2\dot{\bar p}^2+\frac{\epsilon}{8\alpha}(1+\omega){\rho}
\nonumber\\
\Longleftrightarrow
\ddot{\bar p}=-\partial_{ \bar p}{V}(\bar{p},{\rho})-3\epsilon
\bar{p}^2\dot{\bar p},
\end{eqnarray}
where
${V}(\bar{p},{\rho})=-\frac{1}{8\alpha}\left(\bar{p}^2+\frac{{\rho}}{
3\bar{p}^2}\right)$. Then, since ${V}(0,{\rho})=+\infty$ this means that the universe cannot bounce from one phase to another.

Once we have proved that the universe cannot bounce, the next step is to  look for singular solution in the expanding phase and compare them with
the classical ones (equation (\ref{a1})). To do this, we look for solutions with the following behavior at late times
$\rho(t)\sim \rho_0(t_s-t)^{-\nu}$ with $\nu>0$. Inserting this solution in the conservation equation one obtains
$H(t)\sim \frac{\nu}{3(1+\omega)(t_s-t)}$, and finally, inserting both expressions in the modified Friedmann equation (equation (\ref{19})) and
retaining the leading terms, one gets $\nu=-4$ and $\rho_0=-\frac{18\alpha(3\omega-5)}{3(\omega+1)^3}>0$. Thus,
\begin{eqnarray}\label{22}
 H(t)\sim -\frac{4}{3(\omega+1)(t_s-t)},\quad \rho(t)\sim \frac{\rho_0}{(t_s-t)^4}.
\end{eqnarray}

Comparing this solution with (\ref{a1}) we deduce that, in that case, modified gravity make worse the singularities.

To end this Section, note that for more complicated functions $f(R)$, it seems impossible to perform a qualitative analysis of the dynamics. In such cases  
only numerical simulations
could show the behavior of our universe at late times.

\vspace{0.5cm}

\noindent{\it 4.- Loop quantum cosmology}----An approach to quantum cosmology that could  avoid the big rip singularity is
 loop quantum cosmology, where for states which correspond to a macroscopic universe, such as ours,
at late times
the following effective Hamiltonian, which captures the underlying loop quantum dynamics, is considered \cite{as11,s09,s09a}
\begin{eqnarray}\label{4}
{\mathcal H}_{LQC}=-3V\frac{\sin^2( \lambda \beta)}{\gamma^2\lambda^2}+ V\rho ,
\end{eqnarray}
where $\gamma\cong 0.2375$ is the Barbero-Immirzi parameter \cite{m04} and  $\lambda$ is a
parameter with dimensions of length, which is determined invoking
the quantum nature of the geometry, that is, identifying its square with the
minimum eigenvalue of the area operator in LQG, which gives as a result $\lambda\equiv
\sqrt{\frac{\sqrt{3}}{4}\gamma}$ (see \cite{s09a}).
In (\ref{4}), $V$ is the physical
volume $V=a^3$ and $\beta$ is canonically conjugate to $V$ and satisfies $\{\beta,V\}=\frac{\gamma}{2}$,
where $\{,\}$ is the Poisson bracket, which for the canonically conjugate variables $(a,p=-6Ha^2)$ takes the form
$\{\beta,V\}=\frac{\partial \beta}{\partial a}\frac{\partial V}{\partial p}-
\frac{\partial \beta}{\partial p}\frac{\partial V}{\partial a}=-3a^2\frac{\partial \beta}{\partial p}$. Then, since 
$\{\beta,V\}=\frac{\gamma}{2}$ one can conclude that $\beta=\gamma H$.

\begin{remark}
 This last statement does not seems clair in some papers of loop quantum cosmology. For example, in  \cite{cs09} in order to define $\beta$ the 
authors assert that ``On the classical solution $\beta$ is related to the scalar factor as $\beta=\gamma H$'' and in \cite{mp10} it is stated
that `` The variable $\beta$, in the limit $\beta\rightarrow 0$, is linked to the Hubble factor via the relation  $\beta=\gamma H$''.
We will discuss later what really happens with the relation between $\beta$ and $H$.
\end{remark}

The Hamiltonian constraint is then given by 
$\frac{\sin^2( \lambda\beta)}{\gamma^2\lambda^2}=\frac{\rho}{3}$, and the  Hamiltonian equation gives the following identity:
\begin{eqnarray}\label{d}
\dot{V}=\{V,{\mathcal H}_{LQC}\}=-\frac{\gamma}{2}\frac{\partial{\mathcal H}_{LQC}}{\partial\beta}\Longleftrightarrow 
\nonumber\\ H= \frac{\sin(2\lambda \beta)}{2\gamma\lambda}\Longleftrightarrow \beta=
\frac{1}{2\lambda}\arcsin(2\lambda\gamma H).
\end{eqnarray}

Writing this last equation as follows $H^2=\frac{\sin^2(\lambda \beta)}{\gamma^2\lambda^2}(1-\sin^2( \lambda \beta))$ and
using the Hamiltonian constraint ${\mathcal H}_{LQC}=0\Longleftrightarrow \frac{\sin^2( \lambda \beta)}{\gamma^2\lambda^2}=\frac{\rho}{3}$
one obtains the following modified Friedmann equation in loop quantum cosmology
\begin{eqnarray}\label{6}
H^2=\frac{\rho}{3}\left(1-\frac{\rho}{\rho_c}\right)
\Longleftrightarrow \frac{H^2}{\rho_c/12}+\frac{(\rho-\frac{\rho_c}{2})^2}{\rho_c^2/4}=1,
\end{eqnarray}
being $\rho_c\equiv \frac{3}{\gamma^2\lambda^2}$.
This equation with the conservation equation $\dot{\rho}=-3H(\rho+P)$ gives  the dynamics of the universe in loop quantum cosmology.

Here two remark are in order: \begin{enumerate}
                                \item                               
The Hamiltonian (\ref{4}) can be actually constructed by using the general formulae of loop gravity that express  the Hamiltonian
in terms of holonomies $ h_j(\lambda)\equiv
e^{-i\frac{\lambda\beta}{2}\sigma_j} $ where $\sigma_j$ are the Pauli matrices \cite{abl03,t01,aps06}:
\begin{eqnarray}\label{d1} 
&& \hspace*{-5mm} 
{\mathcal H}_{LQG}\equiv-\frac{2 V}{\gamma^3 \lambda^3}
\sum_{i,j,k}\varepsilon^{ijk} Tr\left[
h_i(\lambda)h_j(\lambda)h_i^{-1}(\lambda) 
\right. \nonumber \\ && \left. \times
 h_j^{-1}(\lambda)h_k(\lambda)\{h_k^{-1}(\lambda),V\}\right]
+\rho V.
\end{eqnarray}

%where $\{,\}$ denotes the Poisson bracket
A simple calculation shows, see for instance \cite{he10,dmw09}, that (\ref{d1}) equals (\ref{4}).
\item
The old quantization of loop quantum cosmology 
was done using two canonically conjugate variables, one of them was
the dynamical part of the connection, namely ${\mathfrak c}$, and the other one was
the dynamical part of the triad, namely ${\mathfrak p}$, (see for instance \cite{b02,bv03,cm10}). These variables are related with the scalar factor and
the extrinsic curvature $K=\frac{1}{2}\dot{a}$ by the relations 
%(see Section II of \cite{bv03} and Section II of \cite{cm10})
\begin{eqnarray}
 {\mathfrak p}=a^2, \qquad {\mathfrak c}=\gamma K=\frac{\gamma}{2}\dot{a}.
\end{eqnarray}

Then, in order to obtain the dynamics of the universe,
the following effective Hamiltonian was used \cite{s06,b09,sv05}
\begin{eqnarray}\label{7}
{{\mathcal H}}_{OLC}\equiv -\frac{3}{\gamma^2\mu^2}{\mathfrak p}^{1/2}\sin^2(2\mu {\mathfrak c}) +\rho {\mathfrak p}^{3/2},
\end{eqnarray}
where $\mu=\frac{3\sqrt{3}}{2}$ (see \cite{aps06}) is obtained by identifying the eigenvalue $\frac{\gamma\mu}{6}$ of 
the operator $\widehat{\mathfrak p}$
with the minimum eigenvalue of the
area operator in loop quantum gravity which is given by $\frac{\sqrt{3}}{4}\gamma$.

Using this Hamiltonian, the scalar factor satisfies the dynamical equation
\begin{eqnarray}\label{8}
\dot{a}=\{a,{{\mathcal H}}_{OLC}\}=\frac{\sin(4\mu {\mathfrak c})}{2\mu\gamma}.
\end{eqnarray}
and, imposing once again the Hamiltonian constraint, ${\mathcal H}_{OLC}=0$, one obtains
the following modified Friedmann equation
\begin{eqnarray}\label{9}
H^2=\frac{\rho}{3}\left(1-
\frac{\rho(a)}{\rho_{c}(a)}\right),
\end{eqnarray}
with  critical density \begin{eqnarray}\label{10} \rho_c(a)=\frac{3}{\gamma^2\mu^2a^2}.
\end{eqnarray}

%\item
%If one consider the following modified Hamiltonian
%\begin{eqnarray}\label{11}
%{\mathcal H}_{BC}=-3a^3\frac{\sinh^2(\gamma \lambda H)}{\gamma^2\lambda^2}+ \rho a^3,
%\end{eqnarray}
%the first Hamilton equation gives
%\begin{eqnarray}
%\label{12}\dot{a}=\frac{\partial {\mathcal H}_{BC}}{\partial p}= a\frac{\sinh(2\gamma \lambda H)}{2\gamma\lambda}
%\nonumber\\
%\Longleftrightarrow
%H= \frac{\sinh(2\gamma \lambda H)}{2\gamma\lambda}.
%\end{eqnarray}

%And from this equation and the Hamiltonian constraint one obtains the following modified Friedmann equation
%\begin{eqnarray}\label{13}H^2=\frac{\rho}{3}\left(1+\frac{\lambda^2\rho}{3}\right),\end{eqnarray}
%that could be identified with the modified Friedmann equation in the Randall-Sundrum scenario of brane cosmology \cite{bdel00,cs01} choosing
%$\lambda =\sqrt{\frac{3}{2\rho_{\Lambda}}}$ (being $\rho_{\Lambda}$ the intrinsic tension of the brane).

\end{enumerate}

\vspace{0.5cm}

Coming back to the modified Friedmann equation in loop cosmology, (equation (\ref{6})), from the equation of the ellipsis one can see that
the Hubble parameter belong in the interval $[-\rho_c/12,\rho_c/12]$, and the energy density  $\rho$ in
$[0,\rho_c]$. Then, if the state equation $P=f(\rho)$ is smooth enough, the functions $H(t)$ and $\rho(t)$ will be smooth
and they also will be defined for all
time, that is, there won't singularities.

As an example, we consider the solvable case $P=\omega\rho$. The dynamical equations are now
\begin{eqnarray}
 H^2=\frac{\rho}{3}\left(1-\frac{\rho}{\rho_c}\right),
\qquad\dot{\rho}=-3H(1+\omega)\rho, 
\end{eqnarray}
which solution is given by
\begin{eqnarray}\label{14}
 \rho(t)=\left(\frac{3}{4}(1+\omega)^2(\bar{t}-t)^2+\frac{1}{\rho_c} \right)^{-1},\qquad H(t)=\frac{1+\omega}{2}(t-\bar{t})\rho(t)
\end{eqnarray}
where $\bar{t}=t_0-\frac{2\sqrt{1-\frac{\rho_0}{\rho_c}}}{\sqrt{3}(1+\omega)\sqrt{\rho_0}}$ and $\rho_0$ is the current energy density of our universe.
Note that this solution is defined for all time and it  finishes at $(H=0,\rho=0)$, and its
 main property  is that
the universe  remains in the  expanding phase until time $t=\bar{t}$. At this time it bounces  and re-collapses forever and ever.

%Note also that,
% in brane cosmology, using equation (\ref{13}) one obtains
%\begin{eqnarray}\label{a2}
% \rho(t)=\left(\frac{3}{4}(1+\omega)^2(\bar{t}-t)^2-\frac{1}{2\rho_{\Lambda}} \right)^{-1},\quad
 %\nonumber \\
 %H(t)=\frac{1+\omega}{2}(t-\bar{t})\rho(t),
%\end{eqnarray}
%with $\bar{t}=t_0-\frac{2\sqrt{1-\frac{\rho_0}{2\rho_{\Lambda}}}}{\sqrt{3}(1+\omega)\sqrt{\rho_0}}$. Which shows that, for $\omega<-1$,  there is
%a future singularity at time $t_s=\bar{t}+\frac{1}{\omega+1}\sqrt{\frac{2}{3\rho_{\Lambda}}}$.

It is clear that
the behavior described by equation (\ref{14}), is very different to the classical one (\ref{a1}), where the
universe presents a big rip singularity.

From this result, it seems  that holonomy corrections  replace the big rip singularity, which appears in classical cosmology, by a non-singular bounce.
 However, from our viewpoint, we have some objections to  the way that the modified Friedmann equation has been obtained. 

\begin{enumerate}
 \item 
Loop quantum cosmology was built using two canonically conjugate variables, one is
the dynamical part of the connection, namely ${\mathfrak c}$, and the other one is 
the dynamical part of the triad, namely ${\mathfrak p}$, (see for instance \cite{b02,bv03,cm10}). These variables are related with the scalar factor and
the extrinsic curvature $K=\frac{1}{2}\dot{a}$ by the relations (see Section II of \cite{bv03} and Section II of \cite{cm10})
\begin{eqnarray}
 {\mathfrak p}=a^2, \qquad {\mathfrak c}=\gamma K=\frac{\gamma}{2}\dot{a}.
\end{eqnarray}

Later in \cite{acs08} two new canonically conjugate variables were introduced, $(V,\beta)$ which are related with the standard variables through
the relations $V=a^3$ and $\beta=\gamma H$ (see formulas 2.1 and 2.2 of \cite{acs08}). Then, the loop quantum theory built with this two variables 
provides the effective Hamiltonian (\ref{4}). However, 
if one starts directly from the effective Hamiltonian (\ref{4}),
although it is assumed that $V=a^3$, the definition of the variable 
$\beta$ comes from the equation $\dot{V}=\{V,{\mathcal H}_{LQC}\}$, which gives
$\beta=
\frac{1}{2\lambda}\arcsin(2\lambda\gamma H)$ and differs from the initial definition of $\beta$.
As a consequence, if one takes $V=a^3$ and $\beta=
\frac{1}{2\lambda}\arcsin(2\lambda\gamma H)$ as a canonically conjugate variables, then the standard variables $(a, p=-6Ha^2)$ do not remain
canonically conjugate, because now
\begin{eqnarray}
 \{a,p\}\equiv \frac{\gamma}{2}\left(\frac{\partial a}{\partial \beta}\frac{\partial p}{\partial V}-
\frac{\partial a}{\partial V}\frac{\partial p}{\partial \beta}\right)=\nonumber\\ 
\cos(2\lambda\beta)=\sqrt{1-4\gamma^2\lambda^2H^2}\not=\mbox{constant}.
\end{eqnarray}

\item
One of the main  reason against the modified Friedmann equation in loop quantum cosmology comes from 
  the Legendre transformation 
\begin{eqnarray}
{\mathcal H}_{LQC}=-\frac{2}{\gamma}\dot{V}\beta-{\mathcal L}_{LQC}\end{eqnarray}
which gives, in terms of the standard
variables, the following Lagrangian 
\begin{eqnarray}\label{l1}
 {\mathcal L}_{LQC}=-\frac{3a^3H}{\gamma\lambda}\arcsin(2\lambda\gamma H)+\nonumber\\ 
\frac{3a^3}{2\gamma^2\lambda^2}\left(1-\sqrt{1-4\gamma^2\lambda^2H^2}
\right)-a^3\rho,
\end{eqnarray}
which coincides with ${\mathcal L}_{E}$ for small values of $H$.

It's well-known that the other current cosmological theories are built from two invariant, the scalar curvature $R=6\left(\dot{H}+2H^2\right)$  
and the
Gauss-Bonnet curvature invariant
$G=24H^2\left(\dot{H}+H^2\right)$. For example,  in modified Gauss-Bonet gravity \cite{cenoz06} 
the Lagrangian ${\mathcal L}_{GB}=a^3f(R,G)-a^3\rho$ is used,
 and semiclassical gravity, when one takes into account the quantum effects due to a massless conformally coupled field (see for instance \cite{hae11}),
is based in the trace anomaly $T_{vac}= \alpha\Box  R-\frac{\beta}{2}G$ (being $\alpha>0$ and $\beta<0$ two renormalization coefficients). However,
the Lagrangian (\ref{l1}) does not seem invariant, which is in disagreement with one of the main principles of general relativity.

\item
In semiclassical gravity or in the $f(R,G)=\frac{R}{2}-\frac{\alpha}{12}R^2$ theory, the modified Friedmann equation is given
by 
\begin{eqnarray}\hspace*{-10cm} 
H^2=\frac{\rho}{3}+2\alpha(3H^2\dot{H}+H\ddot{H}-\frac{1}{2}\dot{H}^2)-\beta H^4,\nonumber \\ (\beta=0 \mbox{ in the }
f(R,G)=\frac{R}{2}-\frac{\alpha}{12}R^2 \mbox{ theory }),
\end{eqnarray}
which contains higher-curvature terms like $\dot{H}$ and $\ddot{H}$.
Using this equation, it was proved in \cite{hae11} for the state equation $P=\omega \rho$ that, if $-1\leq \frac{\beta}{3\alpha}\leq 0$ the universe 
will bounce but when it will enter 
in the contracting phase it will develop a future 
singularity of the form \begin{eqnarray}\label{dd}
 H(t)\sim \frac{3\alpha}{\beta}\left(-1\pm\sqrt{1+\frac{\beta}{3\alpha}}\right)\frac{1}{(t_s-t)},\quad \rho(t)\sim 0, \quad \mbox{for} \quad t<t_s
\end{eqnarray}
 being $t_s$ the time at which the future singularity appears. 
%And for, $\frac{\beta}{3\alpha}<-1$ it bounces infinitely many times.
It is clear that this behavior is very different to the one described in loop
quantum cosmology (eq. (\ref{14})), where the universe does not develop any kind of singularity.
\end{enumerate}

\vspace{0.5cm}

On the other hand, one can adopt another different point of view. One can assume that the variables $(a,p=-6H a^2)$ are canonically 
conjugate, which means
that $V=a^3$ and  
$\beta=\gamma H$ as we have seen at the beginning of this Section. Then, one has to understand the Hamiltoninan (\ref{4}), not like the 
Hamilonian of the system, but  as the new Hamiltonian
constraint that replaces the classical one. From this viewpoint,
 taking the derivative with respect to the time of the Hamiltonian constraint 
and finally
using the conservation equation,  the following modified Raychaudury equation will be obtained
%Taking the derivative of the Hamiltonian constrain, which is equivalent to the second Hamilton equation 
%$\dot{\beta}=\{\beta,{\mathcal H}_{LQC} \}$, and using the conservation equation one gets
%\begin{eqnarray}\label{d1}
%\dot{\beta}=-\frac{3}{2}(1+\omega)H\frac{\tan( \lambda \beta)}{\lambda}.
%\end{eqnarray}
\begin{eqnarray}\label{z}
\dot{H}=-\frac{\lambda\gamma H}{\sin(2\lambda\gamma H)}
\left[\frac{3\sin^2(\lambda\gamma H)}{\lambda^2\gamma^2}+f\left( \frac{3\sin^2(\lambda\gamma H)}{\lambda^2\gamma^2}\right) \right].
\end{eqnarray}

For the  case $P=f(\rho)=\omega\rho$ this equation becomes
\begin{eqnarray}
\dot{H}=-\frac{3}{2}(1+\omega)H\frac{\tan(\gamma \lambda H)}{\gamma\lambda},
\end{eqnarray}
from which one deduces that $\dot{H}$ is positive, which means that $H$ reach the value $\frac{\pi}{2\gamma\lambda}$ in a finite time
and thus, at that time $\dot{H}$ diverges or equivalently the scalar curvature $R=6(\dot{H}+2H^2)$ diverges. Moreover, at that time, from the
Hamiltonian constraint  and the state equation $P=\omega\rho$, one has
$\rho=\frac{3}{\gamma^2\lambda^2}$ and $P=\frac{3\omega}{\gamma^2\lambda^2}$.  Then, one can conclude that, from this viewpoint, the
big rip singularity is replaced by this other singularity characterized by a divergent scalar cuvature, but with finite values of the Hubble parameter,
energy density and pressure.

\vspace{0.5cm}

\noindent{\it 5.- Theory based on the reduced semiclassical Friedmann equation}---
This approach was proposed by Parker and Simon in \cite{s92,ps93}, its main idea is to obtain the derivatives of $H$ from the classical Friedmann  and
conservation equations. Thus,  once this has been done,  one inserts these derivatives into the semiclassical Friedmann equation obtained
in modified gravity
(equation (\ref{18}). The result is a new modified Friedmann equation, but without derivatives on $H$.
To be precise, we consider the theory $f(R)=\frac{R}{2}-\frac{\alpha}{12}R^2$. Then,
for the simplest case $P=\omega \rho$ we have $\dot{H}=-\frac{3}{2}(1+\omega)H^2$ and $\ddot{H}=\frac{9}{2}(1+\omega)^2H^3$,
that once introduced in (\ref{19}) provides
\begin{eqnarray}\label{23}
 H^2=\frac{\rho}{3}+\frac{9}{4}\alpha (1+\omega)(3\omega-1)H^4.
\end{eqnarray}

If in this equation one makes the substitution $H^4=\frac{\rho^2}{9}$ (classical Friedmann equation) one will get the interesting equation
\begin{eqnarray}\label{24}
 H^2=\frac{\rho}{3}+\frac{1}{4}\alpha (1+\omega)(3\omega-1)\rho^2,
\end{eqnarray}
which gives a way to obtain the modified Friedmann equations in loop quantum cosmology. Effectively, choosing
$\alpha=-\frac{4\rho_c(1+\omega)(3\omega-1)}{3}$  one obtains the equation
(\ref{6}). Then, in some sense, for the particular state equation $P=\omega \rho$, this approach could justify the modified
Friedmann equation in loop quantum cosmology. However,
note that, for a more general state equation $P=f(\rho)$, this method provides more complicated equations that (\ref{24}), and thus,  it is 
impossible to
 recover equation (\ref{6}).

\vspace{0.5 cm}

\noindent{\it 6.- Conclusions}---
Through this paper we have shown that it seems impossible to avoid completely  the future singularities, which appear in phantom dark energy models,
 using  different alternative approaches to
quantum cosmology. We think the most efficient approach is semiclassical or equivalently modified gravity with $f(R)= \frac{R}{2}-\frac{\alpha}{12}R^2$
(with $\alpha>0$). In this
theory the dynamics of the universe is drastically changed because the universe will bounce and will enter in the contracting phase where it will
develop a future singularity like that described by equation (\ref{20}) or (\ref{dd}).  The other important conclusion in the paper, is 
that the results about
avoidance of future singularities, in particular the bounces, obtained from the modified Friedmann equation in loop quantum cosmology 
(for example \cite{sst06,nw07,sg07})
have to be 
revisited because this equation would not be justified. This does not mean that,
in a more general theory where higher-curvature terms would be combined with holonomy corrections, features such as bounces may appear.
But this is a complicated problem that deserves future investigations.

\vspace{.5 cm}

\noindent{\bf Acknowledgments.} I would like to thak Prof. Martin Bojowald for your critical comments and suggestions about the first version of 
this paper which have been very important in order to improve it.
This investigation has been
supported in part by MICINN (Spain), project MTM2011-27739-C04-01,
%FIS2006-02842 and FIS2010-15640, by CPAN Consolider Ingenio Project,
and by AGAUR (Generalitat de Ca\-ta\-lu\-nya), contract 2009SGR-345.
%and 2009SGR-994.

\end{document}